\documentclass[12pt]{article}

\usepackage{scicite}
\usepackage{epsf}

\usepackage{ulem}
\usepackage{color}

\usepackage[squaren,Gray]{SIunits}
\usepackage{times}
\usepackage{amsmath}
\usepackage{amsfonts}
\usepackage{amssymb}
\usepackage{graphicx}
\usepackage{color}
\usepackage{hyperref}
\usepackage{chngcntr}

\usepackage[utf8]{inputenc}
\usepackage[english]{babel}
\usepackage[table,xcdraw]{xcolor}
\usepackage{amsmath,amsfonts,amssymb}
\usepackage[letterpaper,margin=1in]{geometry}
\usepackage{graphicx}
\usepackage{color}
\usepackage{xcolor}
\usepackage{empheq}
\usepackage{hyperref}
\usepackage{bm}
\usepackage{url}
\usepackage{diagbox}
\usepackage{array}
\usepackage{setspace}
\usepackage{multirow}

\usepackage{soul}

\makeatletter
\renewcommand{\fnum@figure}{\textbf{Figure \thefigure}}
\renewcommand{\fnum@table}{\textbf{Table \thetable}}
\makeatother

\newcommand*{\bapt}[1]{{#1}}
\newcommand*{\baptdel}[1]{}

\newenvironment{sciabstract}{%
	\begin{quote} \bf}
	{\end{quote}}

\def\scititle{
	Unveiling magma plumbing systems for volcanic eruptions and crustal accretion via active-seismic matrix imaging
}

\title{\bfseries \boldmath \scititle}

\author{Baptiste Hériard-Dubreuil${}^{1}{}^\ast$,
	Milena Marjanovi\'c${}^{2, 3}{}^\ast$,
	Arnaud Burtin${}^{2}$,\and
	and Alexandre Aubry${}^{1}$\and\and
	\small$^{1}$Institut Langevin, ESPCI, PSL University, CNRS, Paris, France.\and
	\small$^2$ Institut de Physique du Globe de Paris, Université Paris Cité, CNRS UMR 7154, Paris 75005, France.\and
	\small$^3$ Institute of Geophysics and Planetary Physics, Scripps Institution of Oceanography,\and
	\small University of California, San Diego, La Jolla, CA 92037\and
	\small$^\ast$Corresponding authors; E-mail: baptiste.heriard-dubreuil@espci.fr, marjanovic@ipgp.fr.}

	\date{}
	
\begin{document}
		\baselineskip24pt
		\maketitle
		\begin{sciabstract}
			Submarine eruptions, accounting for over 80\% of Earth’s volcanic activity, primarily occur along mid-ocean ridges, where shallow magmatic systems are accessible to high-resolution imaging. Yet, their remoteness often leaves them undetected. Recent seismic studies at the East Pacific Rise (EPR) 9°50'N-one of the most dynamic ridge segments, imaged the detailed architecture of the shallowest magma lens, but no data-constrained model yet explains how magma accumulates, migrates, or triggers eruptions. Similarly, the formation of oceanic crust remains poorly understood. While 2-D seismic data reveal only a few vertically stacked, transient magma lenses, our study applies matrix imaging, a novel technique in controlled-source seismology, to map the inner structure of on- and off-axis magma reservoirs. We uncover a conical on-axis reservoir and interconnected magma-rich zones throughout the crust. Combined with ophiolite evidence, these findings reveal that magma channels dominate the first 3 km for lower crust formation, while in situ crystallization prevails in the final 1 km, resolving a long-standing debate.
		\end{sciabstract}

		\section*{INTRODUCTION}
		
\begin{figure}[ht]
			\centering
			\includegraphics[width=\textwidth]{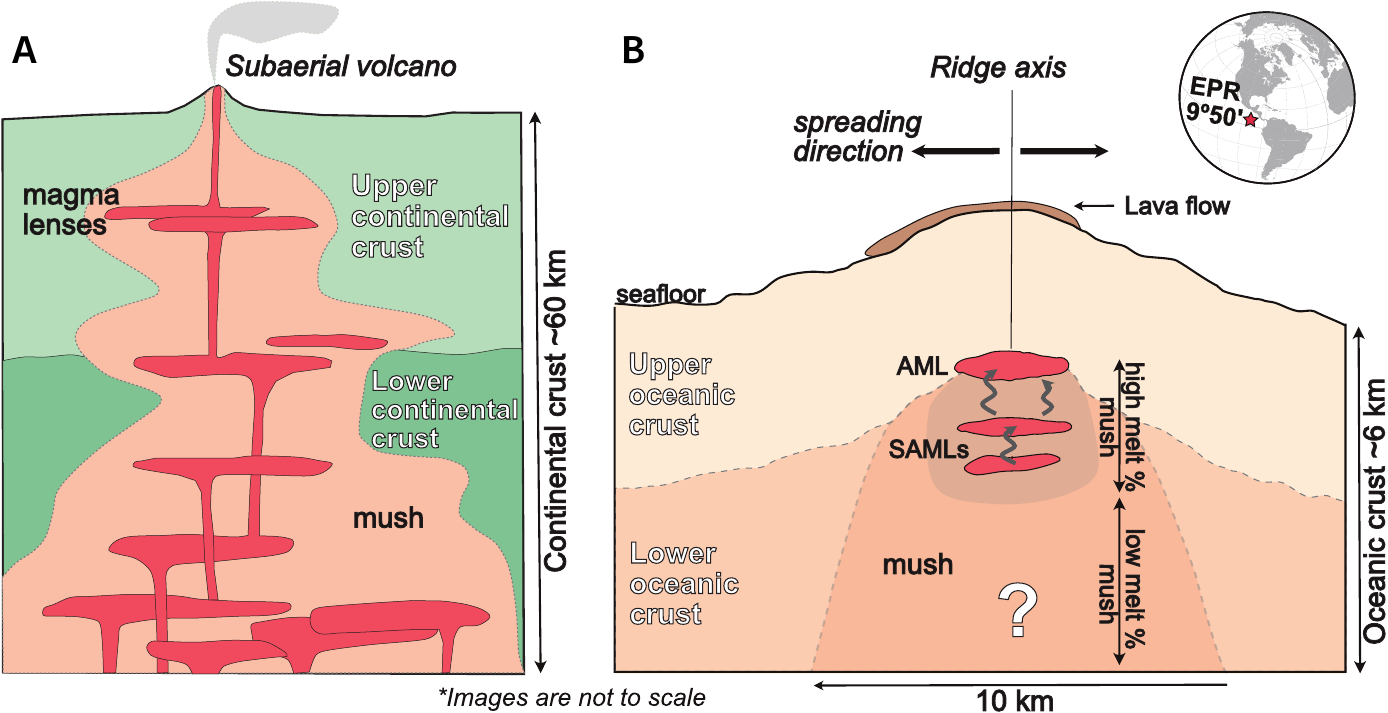}
			\caption{\label{fig1}\textbf{Magma plumbing systems.} (\textbf{A}) \textbf{Subaerial volcanoes:} Conceptual, transcrustal model suggesting the vertical distribution of magma lenses throughout the crustal reservoir. Illustration modified from \cite{cashman2017vertically}. (\textbf{B}) \textbf{Mid-ocean ridges:} Data-based model from seismic observations of the magmatic system beneath the East Pacific Rise (EPR) 9º50’N, indicated on the world globe inset by red star. AML: Axial Magma Lens; SAML: sub-Axial Magma Lens. Illustration modified from \cite{carbotte2021stacked}.
			}
		\end{figure}

		Over a century, our understanding of subaerial volcanic systems has been shaped by conceptual models of magma chambers, ranging from melt-laden, extensive reservoirs \cite{bowen1915later} to transcrustal magmatic systems characterized by vertically stacked lenses within a mush zone \cite{cashman2017vertically} (Fig.~\ref{fig1}A). Based on the latter view, magma reservoirs represent melt-crystal mixtures with localized high melt accumulations ($>30\%$ melt) emerged into mushy, higher crystallinity regions ($<30\%$ melt), delimited by sub-solidus host rock \cite{cooper2017does,sparks2019formation}.
		This model is primarily based on petrological and geochemical analyses of surface-collected samples \cite{solano2012melt, cashman2013volcanoes, annen2015construction, colman2015constraints, cooper2017does, werner2020linking, wieser2020microstructural}. However, reconstructing the processes of melt migration within the crust using petro/geochemical data remains a significant challenge, as surface samples record cumulative effects of chemical interactions from the base of the crust that are difficult to deconvolve \cite{cashman2017vertically}. The most direct evidence for magma reservoirs derives from geophysical data \cite{koulakov2016feeder, johansen2019deep, janiszewski2020aseismic, wawrzyniak2025magnetotelluric}, particularly controlled-source (active) seismic \cite{kiser2016magma, kiser2018focusing, autumn2025exploring}. The largest land-based active seismic survey, conducted at Mount St. Helens in the Cascade arc, produced compressional and shear velocity models and identified a low-velocity anomaly at 4-6 km depth, $\sim$20 km wide, which was interpreted as a magma reservoir with 10-12\% inferred melt content.
		The observed velocity anomaly was corroborated by high-conductivity zones in magnetotelluric images \cite{ulberg2020local}. However, the spatial resolution remained limited to tens of kilometers, restricting detailed imaging \cite{kiser2016magma, kiser2018focusing}. Therefore, understanding the transcrustal magmatic system requires reassessment of processes controlling magma migration, accumulation, and crystallization. 
		As the mid-ocean ridges emerged from the seafloor panorama and a wealth of interdisciplinary data was acquired at sea, it became evident that this $>$65,000-km-long mountain chain stores the most active magmatic apparatus in the solar system. Its dynamics appear to dominate the genesis of $\sim$70\% of Earth’s surface at rates ranging from slow to fast, and govern volcanic eruptions. In some parts of this system, eruptions have been documented on a decadal scale e.g., Axial Volcano and the Endeavour segment along the Juan de Fuca Ridge \cite{weekly2013termination, wilcock2016seismic}; East Pacific Rise at 9º50’N \cite{tolstoy2006sea, tan2016dynamics}. The interior of these submarine magmatic systems is accessible to modern high-resolution seismic techniques. Similar to subaerial (land-based) volcanoes, they were initially  depicted as kilometric magma chambers, large, kilometer-scale pools of molten rock beneath the surface \cite{cann1970new, cann1974model, pallister1981samail, smewing1981mixing}.
		This view persisted until the first image of an axial magma lens (AML; a thin, lens-shaped body of molten rock beneath the axis of a mid-ocean ridge) at the East Pacific Rise (EPR) 9º50’N was obtained \cite{herron1978structure, herron1980magma}. Subsequent imaging of a nearly continuous lens along the axis \cite{detrick1987multi} established a new concept: a mush-dominated reservoir, a zone containing a mixture of melt and solid crystals \cite{detrick1993seismic, dunn2000three}. In this model, the AML sits atop a mushy melt-crystal mixture that extends to the base of the crust (Fig.~\ref{fig1}B). Tomography models suggest that the reservoir's width ranges from 1-2 km at the top (matching the AML) to 8-10 km near Moho (the boundary between Earth's crust and mantle) depths \cite{dunn2000three}. Furthermore, modern seismic exploration has revealed not one, but a couple additional vertically stacked magma lenses beneath the mid-ocean ridge - MOR \cite{marjanovic2014multi,canales2017seismic, marjanovic2018crustal,boddupalli2019distribution, carbotte2020stacked, carbotte2021stacked}. However, except for the Axial Volcano at the ridge-hotspot intersection, all MOR magmatic systems show magma accumulations within the first kilometer beneath the AML (Fig.~\ref{fig1}B).
		
		The role of these stacked sills in the formation of oceanic crust, foremost its lower layer beneath the AML, is a subject of an ongoing debate, with “gabbro-glacier” and “sheeted-sill” representing end-member models. According to the gabbro-glacier model, the lower gabbroic section is formed by the crystallization of magma from a single AML, which is then transported downward and outward as the plates move apart. However, this mechanism is difficult to reconcile with the observation of layered gabbro dominating the lowermost part of the crust exposed in ophiolites, an analogue of oceanic crust, which, due to the obduction process, is exposed on land \cite{nicolas1988new, nicolas2009subsidence}. The opposing mechanism was proposed, suggesting that the crystallization occurs in situ through multiple-level sills. However, the sills extending to the base of the crust have not been observed along typical magmatically spread MOR \cite{carbotte2021stacked}. In addition to the crystallization of melt, the role of these magma lenses in its transportation through the system is elusive, with two predominant views: rapid migration of melt by hydrofracturing channels \cite{kelemen1998periodic, korenaga1998melt} or reactive porous flow \cite{macleod2000fossil,carbotte2020stacked, lissenberg2013pervasive}. Therefore, imaging the lenses and their connectivity has important implications not only for understanding crustal formation but also for the eruption mechanism at MOR and subaerial volcanoes.
		
		Recently, a novel approach, referred to as "matrix imaging", has been proposed for in-depth imaging of volcanoes at a resolution of the order of the wavelength. Originally conceived for medical ultrasonics \cite{lambert2020reflection, bureau2023three} and optical microscopy \cite{yoon2020deep, badon2020distortion}, matrix imaging emerged as a promising tool for penetrating the interior of the subaerial volcanoes using reflected bulk wave components in ambient noise data. A preliminary study concerned the Erebus volcano in Antarctica, where the first implementation of matrix imaging revealed its chimney-shaped structure at shallow depths as well as a magma reservoir and several cavities in depth \cite{blondel2018matrix} A second case study example was focused at La Soufriere volcano in Guadeloupe, where matrix imaging revealed the complex geometry of the magmatic accumulations and the conduit connecting the magmatic system and the volcano crater \cite{giraudat2024matrix}. Another example is a study at  Mount St. Helens, which showed the presence of several small (2 km wide) anomalies associated with magma reservoirs \cite{wei2025multilayered}, instead of a large, tomography-derived, one \cite{kiser2016magma}. However, matrix imaging is, by no means, restricted to passive seismic data. Here, we apply it, for the first time, to active source seismic data to help illuminate the architecture of the magmatic system beneath the EPR 9º50’N. 
		
		The EPR 9º50’N (Figs~\ref{fig1}B; S1) is a unique natural laboratory that represents one of the best studied portions of magmatically spread MOR segments with three documented eruptions in 1991-92, 2005-06 and 2025 \cite{haymon1993volcanic, rubin1994210po, tolstoy2006sea, soule2007new, tan2016dynamics, walterdirect, wosniak2026direct}. Recent groundbreaking results obtained using the 3-D multichannel seismic reflection (MCS) dataset provided the first images of the shallow most magma lens (Fig.~\ref{fig1}B) in ultra-high resolution (25 m by 25 m); these results led to a new understanding of the structure of the EPR magmatic system showing that the top surface of the AMLs is corrugated with elongated ridges and troughs from where repeat dikes nucleate leading to seafloor eruptions \cite{marjanovic2023insights}. Although the group of SAMLs at only a couple of hundred meters below the AML was for the first time reported at the EPR 9º50’N \cite{marjanovic2014multi}, their 3-D geometry has not been mapped. In addition to the ridge-centered magmatic system, the 3-D seismic data revealed the ubiquitous presence of off-axis magma lenses (OAMLs) \cite{canales2012network, han2014architecture, aghaei2017constraints}, their origin and role in crustal processes is poorly understood.
		
		This paper addresses the fundamental challenge of imaging the EPR magmatic system at unprecedented depth and resolution. In contrast with concurrent seismic imaging methods, matrix imaging does not rely on a sophisticated wave velocity mode. At the EPR, the best available velocity model only constrains the velocity of the first kilometer of the upper crust, enabling proper migration of just the topmost magma body, i.e., AML  \cite{marjanovic2018crustal}. This velocity model was obtained using a 3-D elastic full-waveform inversion (FWI) algorithm focused on the turning signal at mid- to far-offsets \cite{marjanovic2018crustal}. Furthermore, the deeper magma bodies (SAMLs) are surrounded by mush and would require application of a 3-D viscoelastic FWI algorithm, associated with an elevated computational cost. On the contrary, matrix imaging relies on physics-based adaptive focusing strategies that can be applied in the post-processing stage through the reflection matrix concept. 
		
		The reflection matrix contains the responses between a set of sources insonifying the medium and a set of receivers acquiring the back-scattered wave-field. Once recorded, this matrix holds all the available information on the underground reflectivity, and a set of operations can be applied to retrieve an image of the subsurface. In particular, a numerical focusing process often referred to as redatuming in seismology can be applied to provide the response between virtual geophones at depth and evaluate the focusing quality everywhere inside the medium \cite{lambert2020reflection, touma2021distortion}
		This focusing quality is a direct indicator of the gap between the FWI velocity model and the real wave velocity distribution inside the Earth’s crust. Inspired by seminal works in astronomy \cite{babcock1953possibility}, optical microscopy \cite{badon2020distortion} and ultrasound imaging \cite{bureau2023three}, an adaptive focusing strategy is then employed to compensate for the phase distortions induced by this mismatch \cite{touma2023imaging}. The result is a 3D reflectivity image of the EPR magmatic system up to a 6 km-depth beneath the sea-floor at a resolution of the order of the wavelength ($\sim$\baptdel{250 m}\bapt{500 m}). 
		
		This image sheds a new light on the magmatic activity at work at the EPR. In particular, it reveals: (i) the on-axis conical shape of the magma reservoir; (ii) the presence of vertically stacked anomalies that extend to the Moho, with different roles in crustal formation and volcanic eruption.

		\section*{RESULTS AND DISCUSSION}

		\subsection*{Matrix Imaging applied to controlled-source seismic data }
		
\begin{figure}[ht]
			\centering
			\includegraphics[width=\textwidth]{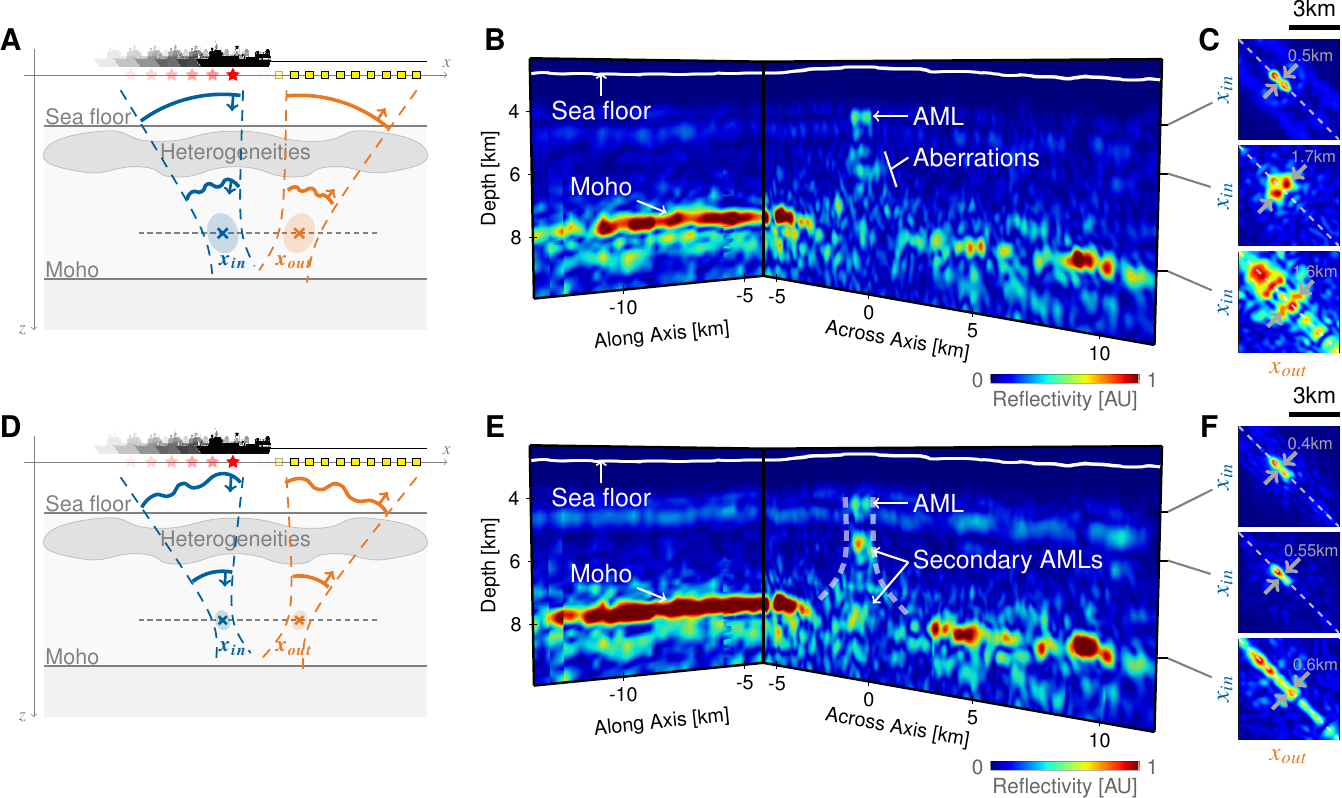}
			\caption{\label{fig2}\textbf{Matrix Imaging applied to active seismic reflection imaging.} (\textbf{A)} Schematic representation of wave focusing in emission (blue) and reception (orange) through an aberrating medium and creating virtual sources and receivers. The fading effect of the ship, sources, and receivers mimics the ship motion. (\textbf{B)} Slice view of the reflectivity volume obtained with matrix imaging. \textbf{C)} Focused reflection matrices at depths $z=$4.1, 5.4 and 8 km, representing the responses between virtual geophones. Arrows indicate the estimated image resolution. (\textbf{D} to \textbf{F}) Same types of images as in panels (A)-(C) but compensating for the aberrations caused by the velocity heterogeneities, leading to sharper virtual geophones and revealing the structure of the axis centered reservoir delimited by white dashed line in panel (B).
			}
		\end{figure}

		Previous studies in seismology applied the matrix imaging tool to ambient noise data \cite{blondel2018matrix, touma2021distortion, giraudat2024matrix, wei2025multilayered}. Here, for the first time, this generic approach is applied to active seismic data (Fig.~\ref{fig2}), combining emissions along the ship path and receptions along streamer lines (Methods) to virtually focus seismic P-waves and synthesize virtual geophones directly inside the medium (Fig.~\ref{fig2}A, Methods). As a wave velocity model, we choose the result of FWI over the first five kilometers beneath the sea surfave with a constant extrapolation for greater depths (\cite{marjanovic2018crustal}; Fig.~\ref{figs4}).
		In contrast to the previous studies that only considered a layered velocity model \cite{touma2023imaging}, the redatuming process is here performed using the split step Fourier method \cite{hardin1973application} that accounts for the lateral velocity variations provided by FWI (Methods).
		The result is a focused reflection matrix $\mathbf{R}_{xx}(z)=[R(x_{in}, x_{out}, z)]$ at each depth $z$ that contains the impulse responses $R(x_{in}, x_{out}, z)$ between virtual geophones acting as sources at $\bm{r}_{in}=(x_{in}, z)$ and others acting as receivers at $\bm{r}_{out}=(x_{out}, z)$ (Fig.~\ref{fig2}C, Methods). On the one hand, each diagonal of these matrices ($x_{in}=x_{out}$) provides a reflectivity image of the subsurface, displayed here as the envelope of the stacked migrated signal (Fig.~\ref{fig2}B). On the other hand, the focusing quality can be assessed everywhere by considering the off-diagonal spreading of energy in $\mathbf{R}_{xx}(z)$ ($x_{in}\neq x_{out}$, Fig.~\ref{fig2}C). 
		
		While the diagonal feature of $\mathbf{R}_{xx}(z)$ indicates a diffraction-limited resolution at shallow depth, the off-diagonal spreading of energy in $\mathbf{R}_{xx}(z)$ at larger depths is the result of wave velocity anomalies that are not grasped by the FWI velocity model (Fig.~\ref{fig2}C). The raw confocal image displayed in Fig.~\ref{fig2}B is therefore polluted by residual wavefront distortions accumulated by bulk seismic waves during their propagation through the Earth crust. This image should thus be interpreted carefully at this stage, especially at large depths. The compensation of such aberrations is then performed through virtual adaptive focusing (Fig.~\ref{fig2}D, Methods). The corrected matrix $\mathbf{R}_{xx}(z)$ now shows a diagonal feature at any depth (Fig.~\ref{fig2}F). A diffraction-limited resolution is therefore reached throughout the volcano volume. The corresponding image displayed in Fig.~\ref{fig2}E provides a reliable representation of the EPR magmatic system over a 10 km-depth range. As we show, this image provides an unprecedented view on the EPR deep magma bodies and contains a wealth of information about their connectivity.
		
		\subsection*{Benchmark for matrix imaging using active seismic}
		
		To validate that matrix imaging applied to active-source seismic data captures real geological events, results from matrix imaging were compared with  images of the AML \cite{marjanovic2023insights} and Moho \cite{aghaei2014crustal}, both obtained using seismic data processing flows. In the matrix imaging process, we used the dominant frequency range of 5 - 15 Hz; as expected, this resulted in lower resolution than obtained by the conventional seismic imaging, but the overall geometry of the events was well recovered. Examining the AML, we observe comparable event widths in the cross-axis direction, as well as undulations and gaps in the reflector (fig. S2). When focusing on the Moho reflector, it is well imaged throughout the volume as a strong, continuous signal extending to $\sim$2 km to the ridge axis (fig. S3). Notably, this imaging gap aligns with observations in the post-stack migrated seismic reflection cube \cite{aghaei2014crustal}. More sophisticated processing has revealed Moho beneath the axis in certain locations, but prominent gaps are still present in particular where the AML is shown to be dominantly melt-enriched. Based on these clear similarities, we attribute the coherent signals observed in confocal images to real geological events, specifically components of the magma plumbing system beneath the EPR at 9º50’N.

		\subsection*{Three-dimensional imaging of distinct levels of magma accumulations within the axial magma reservoir}
		
		The obtained confocal images reveal the presence of multiple anomalies beneath the AML, we interpret them as sub-AML accumulations (SUBA), to make a distinction with the group of reflection events reported in 2-D data interpreted as SAMLs. The SAMLs are reflectors interpreted as individual lenses, whereas SUBA represent anomalous regions that encompass two or more SAMLs. Two SUBA anomalies are particularly prominent as they can be tracked along the ridge axis, sitting at 1-2 km (SUBA1), and 2-3 km (SUBA2) below the main AML (Fig.~\ref{fig3}). The shallower, SUBA1, spans the depth of all previously reported SAML reflectors in 2-D \cite{marjanovic2014multi}. Our 3-D results show that this anomaly follows almost continuously the AML,  and is on average the same size as the AML above (approximately 1km wide). An increased width is observed in the central part spanning latitudes 9º46’N to 9º50’N. 
		
		The second level of deeper accumulation, SUBA2 (Fig.~\ref{fig3}F), was not observed in 2-D conventional seismic imaging. In the confocal volume, it is represented by a wider (from 1.5 to 2.5km wide) and intermittent feature, with three clearly defined magmatic segments disrupted at 9º45-9º46’N and 9º52’-9º53.5’N. Both disruptions are collocated with the prominent tectonic discontinuity identified in the seafloor \cite{carbotte2013fine}, which corroborate the previous view that the larger tectonic discontinuities extend throughout the crust \cite{marjanovic2018crustal}. 
		Below these two clear SUBA regions, we detect a distinct zone extending 3-5 km below AML. It is populated by a large number of smaller randomly distributed anomalies within a zone that is at least twice the width of SUBA1 i.e., $\sim$4 km around the axis. Their average size across axis direction is $<$1 km and in along axis $<$3 km, which are randomly scattered in this lower most part of the crust to Moho, i.e., mantle transition zone (MTZ). It is within this zone on the axis that we also do not observe the anomalies associated with the Moho reflector.

		\subsection*{Implications for the formation of the oceanic crust}
		
		Based on previous observations at EPR 9º50’N and elsewhere, it was concluded that the observed SAML events are ephemeral features, with their presence/absence reflecting the phases of increased/decreased melt delivery to the system \cite{carbotte2021stacked}. The imaging capabilities were considered as an unlikely cause for absence in the seismic reflection section. However, our approach shows that the signal is present in the data. The attenuating, viscoelastic realm of the mushy reservoir is known to shift the dominant frequencies of the propagating signal to lower values \cite{wilcock1995seismic}, which precludes the detection of the deep reflection signal using conventional seismic processing but is enabled via matrix imaging that exploits frequency $\sim$10 Hz. The distinct character of the anomalies we observe argues for differences in their origin and divides the crustal reservoir into two distinct magmatic domains. 
		The upper domain extending to $\sim$3 km below the AML is narrow ($<$2 km) and dominated by vertically stacked anomalies, each encompassing one or more magma lenses. These two SUBA domains match the extent of the lower-velocity region reported by tomography study at 9º30’N, which shows a drop in compressional velocity from $\sim$6 km/s at AML to $\sim$4.2 km/s at SUBA1, then gradual increase to 6 km/s at the depth of SUBA3. The observation that the lowest velocity is centered at the SUBA1 level rather than at the AML supports the idea that the SUBA anomalies, we identify, represent a cluster of several lenses. Based on the temperature model derived from velocity \cite{dunn2022dual}, the melt content within this region is estimated to vary from 6 to 18\%. However, based on the effective medium theory \cite{mainprice1997modelling}, we predict up to 50\% of melt within the Vp minimum.
		
\begin{figure}[ht]
			\centering
			\includegraphics[width=\textwidth]{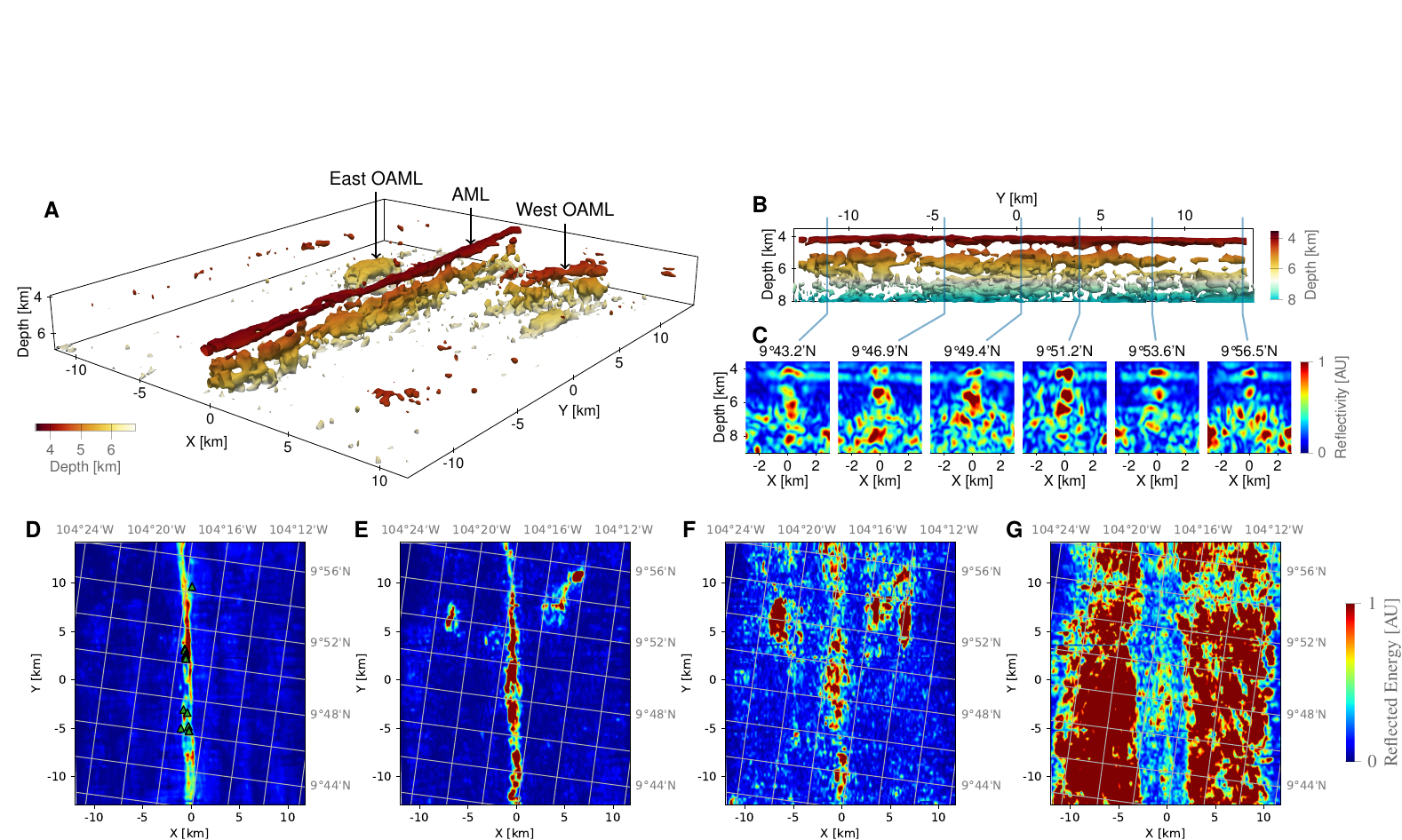}
			\caption{\label{fig3}\textbf{Matrix Imaging reveals multiple magma bodies below the ridge axis.} (\textbf{A}) Perspective view of iso-surfaces obtained with matrix imaging, color coded as a function of depth. We observe multiple bodies below the ridge axis (AML, SUBA1 and SUBA2) and two off-axis magmatic systems. (\textbf{B}) East view of the ridge axis. Three levels of magmatic bodies are present all along the ridge axis. (\textbf{C)} Across-axis slices of the reflectivity image at six different positions, revealing the triangular shaped magma reservoir. (\textbf{D} to \textbf{G}) Top view of the reflected energy integrated for the regions (\textbf{D}) 3.5 - 4.5 km below sea surface (bss) for top-most AML, (\textbf{E}) 5 - 6 km bss indicated as SUBA1, (\textbf{F}) 6 - 7 km bss indicated as SUBA2 and (\textbf{G}) 7 - 9 km bss indicated as SUBA3 from sea surface. Green triangles in D represent hydrothermal vents.
			}
		\end{figure}

		From the geological perspective, this upper zone corresponds to the layer of foliated gabbro in ophiolites \cite{nicolas1988new, boudier1996magma, vantongeren2008cooling, vantongeren2021composition} and exposed oceanic crust (Hess and Pito Deep; \cite{koepke2022reference}). Previous work suggested that the observed SAMLs have also been placed in the foliated section, but with no contribution in their creation. Instead, the nature of lenses was interpreted as transient with a primary role in melt delivery to feed an eruption \cite{carbotte2021stacked}. Based on our results and in concert with the extensive observations from exposed sections \cite{kelemen1997geochemistry}, we argue that these accumulations play an important role in formation of the upper crust through vertical channels connecting the magma lenses, instead of subsidence of crystals and porous flow \cite{macleod2000fossil, lissenberg2004structure, lissenberg2013pervasive}. As these channels are vertical, they are invisible in our data, but the evidence of the channels’ presence is clear in the off-axis magma lenses described in the following section. This, however, does not exclude the possibility that the melt migration between individual cluster of lenses is dominated by porous flow \cite{carbotte2021stacked} and in addition to the mixing of replenished and interstitial melt, lenses may grow by segregation of interstitial melt and compaction \cite{lissenberg2019consequences}.  Finally, our view that these deeper lenses do not represent only passive accumulations for melt delivery to the AML is also supported by recent ultra-high-resolution images of the AML, which show that the eruptions at the EPR 9º50’N are primarily triggered by excess melt channeled from the Lamont mantle melt anomaly \cite{marjanovic2023insights} and are only partially fed from below \cite{marjanovic2014multi}. We also imaged portions of this channel as a group of scattered anomalies on the western flank between 9º50’N and 9º52’N (Fig.~\ref{fig3}E).
		
		The remaining couple of kilometers of the crust, represented by layered gabbro in ophiolites, are characterized by scattered melt accumulations, we place within the Moho transition zone (MTZ), where the composition and texture of the intrusions in dunite were found to be similar to the layered gabbro in ophiolites \cite{kelemen1997geochemistry}. The organization of the melt into these small domains, points to distinct mechanisms for their origin. The anomalies we detect could result from self-organization due to crystal concentration and compaction, as predicted by numerical models \cite{richter1984dynamical}.
		This will lead to segregation of melts, which when large enough will start to move upwards as a porosity wave. The segregation is expected to be more prominent at the mush host-rock contact and is channeled toward the axis resulting in a much narrower zone. One portion of this melt directly beneath the axis is ascending directly to AML via porosity flow, which can explain the compositional (foremost Fe-Mg) equilibrium between the lower crust and intruded/extruded section in the upper crust. However, some of the melt, pushed against the shoulders of the MTZ, is channeled through dikes and accumulated within SUBA1 and SUBA2, which then build the foliated section. The change in width of the crustal reservoir we observe between these distinct domains could have an important impact on the redistribution of the anomalies and thereof, in magma plumbing. The small dimensions, and distribution in small randomly distributed anomalies offer an explanation for the concentration of the velocity anomaly only in the upper part reported by tomography models \cite{dunn2022dual}. 
		
		\begin{figure}[ht]
			\centering
			\includegraphics[width=\textwidth]{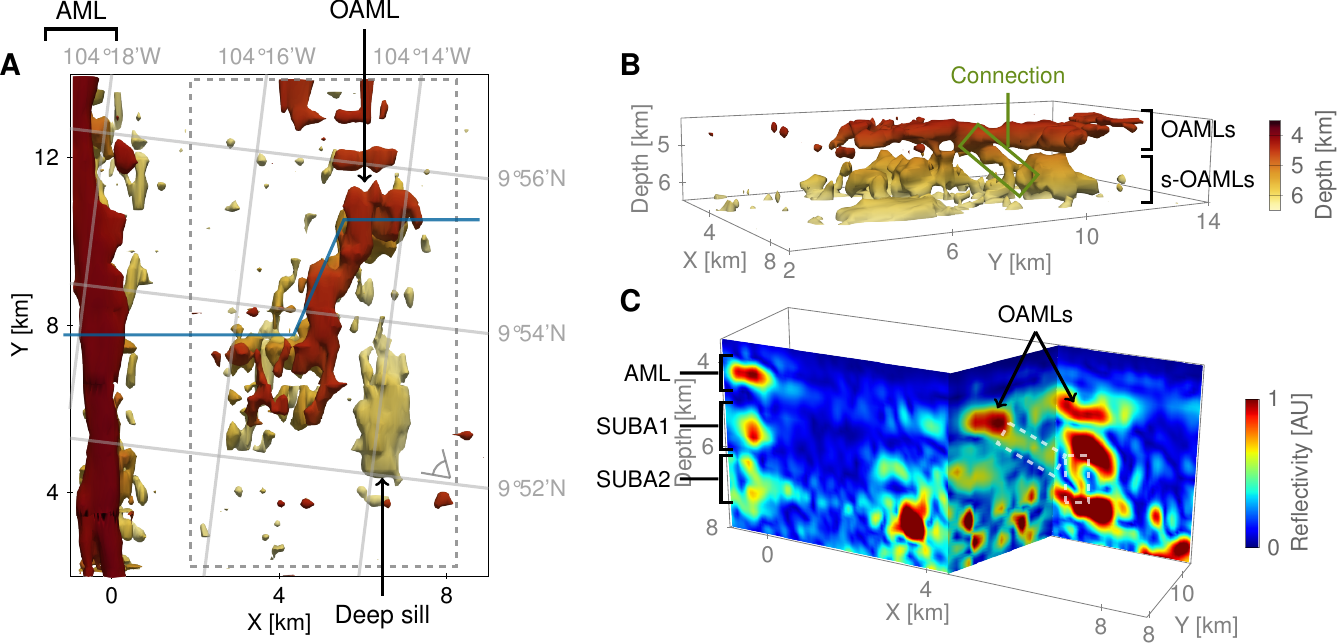}
			\caption{\label{fig4}\textbf{ Imaging magma pathways.} (\textbf{A}) Plan view of the magmatic system in the NW part of the survey area that encompasses the AML and complex off-axis magma lens (OAML). The dashed rectangle is represented in b) from the point of view indicated in gray. (\textbf{B}) Zoomed in portion of panel (A) seen from south east. OAMLs and sub-OAMLs can be identified, with potential connections in green. (\textbf{C}) Fence diagram extracted from the reflectivity volume. Similarly to the on-axis magmatic system, the off-axis system exhibits vertically stacked organization of interconnected magma lenses with clearly identified pathways (white dashed rectangles) that had only been speculated to exist so far \cite{cashman2017vertically}.
			}
		\end{figure}
		
		\subsection*{Multiple {off-axis} magma bodies and implications from subaerial volcanoes}
		
		In addition to the on-axis intra-crustal magma reservoir beneath the axis, we also image vertically stacked lens organization of the magma reservoir beneath the OAMLs. The setting in which they develop is more similar to a subaerial volcano than the reservoir beneath the ridge axis where extensional stresses dominate. The fact that the multiple level anomalies develop even beneath these OAMLs suggest that this must be the uniform mode that exists across different volcanic systems. 
		
		The best developed of these is the system of OAMLs on the eastern flank for which only the topmost lens was previously reported in time-migrated seismic volume \cite{canales2012network}. The anomalies beneath this OAML are distributed in a broader reservoir, extending approximately 5 x 7 km laterally. Only the topmost OAML forms a single anomaly, whereas at the deeper levels that anomalies appear in patches variable in size, from 500 m (imaging resolution) to 3 km along and across the ridge-axis direction. Another significant observation is the presence of signals that interconnect them, which we interpret as the magma pathways through which magma is mobilized through the reservoir. The fact that we image these pathways argues that magma is mobilized by systems of dikes, rather than as a porous flow, which is proposed to dominate the on-axis magma movement \cite{carbotte2020stacked,carbotte2021stacked}. 
		
		In our images only the pathways with an inclination could be imaged, but that does not exclude the existence of the vertical ones. Another important difference is that the composition of the expelled lavas above them does not reflect only the composition of the mantle source beneath them, but instead it must be convolved with the signal of the recrystallized lower crust created on the axis. In previous work it was speculated that the topmost OAML may connect to the main magma body (i.e., AML), but, in the confocal volume, this is not visible. The presence of wehrlite was reported in ophiolites and they were interpreted to form $>$1 km off-axis in already crystallized host rock \cite{quick1993ductile}. They were explained by the dragging of the lenses from the ridge axis or by thermal process. Therefore, these OAML reservoirs may not have significant contribution in formation of the crust, but they introduce heterogeneities and represent a paradigm model for the lenses' organization beneath volcanoes. 
		
		\begin{figure}[ht]
			\centering
			\includegraphics[width=\textwidth]{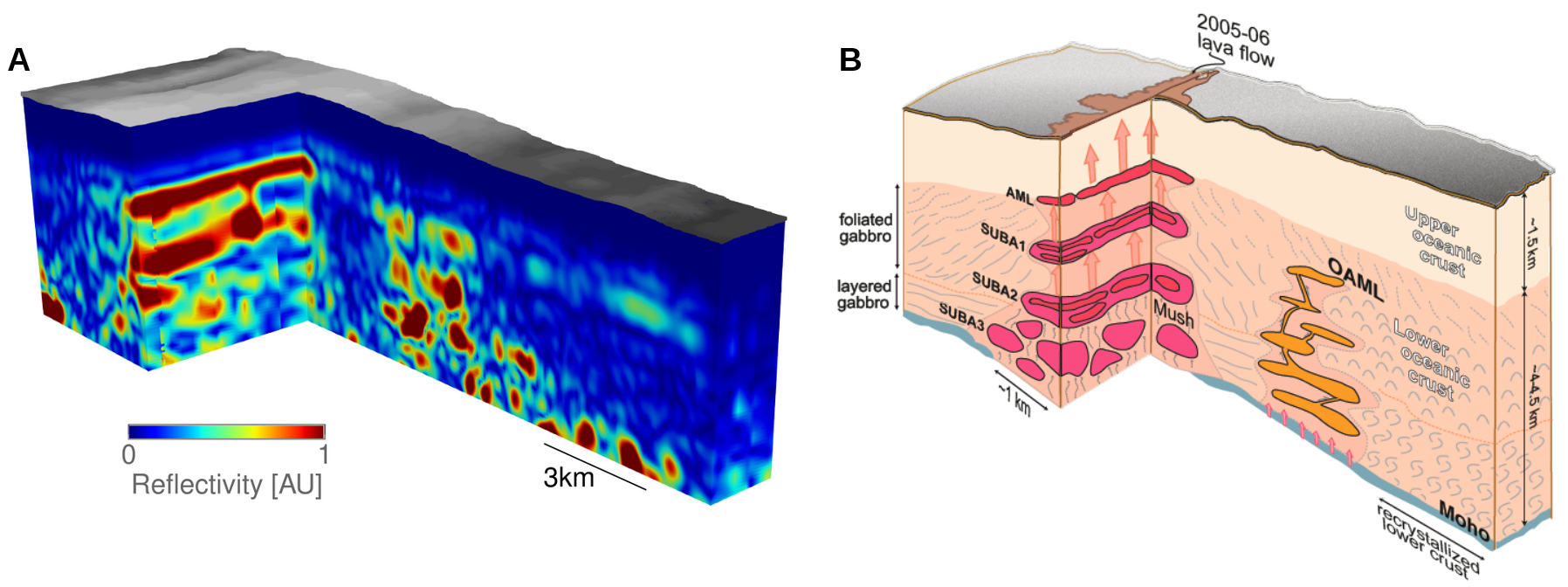}
			\caption{\label{fig5}\textbf{Magma reservoirs as {revealed} by matrix imaging.} (\textbf{A}) Reflectivity volume obtained with matrix imaging. (\textbf{B}) Schematic representation of the on-axis and off-axis magma reservoirs with complex magma accumulations derived from (A).
			}
		\end{figure}
		
		\subsection*{Matrix imaging as a guide star for full waveform inversion}
		
		In this paper, we have fruitfully combined FWI and matrix imaging tools to provide an unprecedented view on the magmatic plumbing system of EPR. Matrix imaging should not be regarded as a substitute for seismic imaging; instead, it should be viewed as an additional tool that can provide complementary insights of the subsurface. Whilst matrix imaging can be performed without prior knowledge of the velocity model, a well-constrained forward model provided by FWI helps significantly in \baptdel{reducing off-diagonal signal in focused reflection matrices}\bapt{resolving sub-AML heterogeneities} (Fig.~\ref{figs4}).  
		
		More generally, we believe that the matrix imaging framework could be advantageously incorporated into a more general optimization scheme to guarantee the convergence of FWI towards a physically-grounded solution. A first step towards this ambitious aim has been recently provided in ultrasound \cite{heriard2026physics}. Using the focusing quality as an intrinsic guide star, a learning-based formulation of matrix imaging has been proposed to iteratively recover the wave velocity landscape from the reflection data. One perspective of this work will be therefore to extend this approach to seismology to recover a high resolution map of the wave velocity inside EPR and other volcanoes, thereby turning matrix imaging into a quantitative tool. The compressional wave velocity can actually be a highly sensitive probe of the magma presence, of its melting state and its gas content \cite{carrara2022dispersive}. The combination of matrix imaging and FWI can therefore become a revolutionary game changer in the way scientists understand and model volcanic systems.

		\section*{MATERIALS AND METHODS}
		
		\subsection*{Data Acquisition}
		
		Data were collected in Summer 2008 aboard R/V Marcus G. Langseth during the MGL0812 3D seismic survey. To acquire the data, four 6-km-long solid streamers, each with 468 hydrophones spaced at 12.5 m, were deployed at 7.5 m below sea surface. To optimize the time for data collection, race track acquisition geometry was adopted, with three race track loops (fig.~S1). The final box was completed with reshot and infill lines. For all lines, the distance between the cables was set to 150 m. Well-tuned, two-array seismic sources with 3300-in3 total volume firing in alternating mode were used to produce acoustic energy at 37.5-m space interval. The maximum recorded time was $\sim$10 s, and the sampling rate was 2 ms.
		
		\subsection*{General Approach}
		
		A rectangular region is selected for this study, spreading between 9º42N and 9º56N and between 104º56W and 104º23W, corresponding to the region with dense acquisition lines (orange rectangle in fig. S1). For each line intersecting this region, all signals recorded by receivers in the selected region and corresponding to an emission in the same region are collected, cleaned, and processed using the matrix imaging method. We use a 2D configuration along lines perpendicular to the ridge. The obtained reflectivity images are stacked along the ridge direction to form a 3D volume of reflectivity, exploited either through slices, iso-surfaces or integrated in sub-volumes.
		
		\subsection*{Data Cleaning and Pre-processing}
		
		To prepare the raw data for the matrix imaging approach, we filter the recorded signals over the 5-15 Hz frequency range. We then proceed to cleaning up the data, removing emissions and receptions of outlier energy. To prevent artifacts caused by the sea-floor reflections, we select the time interval between the first and second sea-floor reflection (excluded). We then perform a temporal Fourier transform of the cleaned signals and stacked them into an experimental reflection matrix, $\mathbf{\widehat{R}}_{uu}(f)=[\widehat{R}(u_{in}, u_{out}, f)]$, for each frequency, whose elements $\widehat{R}(u_{in}, u_{out}, f)$ correspond to the signal recorded at position $u_{out}$ following an emission at position $u_{in}$. This experimental reflection matrix is re-interpolated so that emissions and receptions lie on regularly arranged positions on a line perpendicular to the ridge axis, with a spatial sampling of half the wavelength at the sea surface (75m). Finally, this matrix is symmetrized to compensate for the asymmetry between emissions and receptions, supposing the time reciprocity of the wave phenomena involved.

		\subsection*{Matrix Imaging – Image Formation}
		
		Once the experimental reflection matrix is obtained, our goal is to project it inside the medium to retrieve an image of its  reflectivity. To do so, we adopt the matrix imaging framework, which consists in creating virtual geophones at each depth $z$. Technically, we perform a synthetic focusing operation by multiplying our experimental reflection matrix by the adjoint of the Green’s matrices at all frequencies on both sides \cite{touma2021distortion}:
		\begin{equation}
			\mathbf{\widehat{R}}_{xx}(f, z) = \mathbf{G}_{ux}^\dag(f, z) \times \mathbf{\widehat{R}}_{uu}(f) \times \mathbf{G}^\ast_{ux}(f, z),
		\end{equation}
		where the symbols ${}^\ast$ and ${}^\dag$ stands for phase conjugation and hermitian transpose. $\mathbf{\widehat{R}}_{xx}(f, z)=[\widehat{R}(x_{in}, x_{out}, f, z)]$ is the focused reflection matrix at depth z. $\mathbf{G}_{ux}(f,z)=[G(u_{in}, x_{out}, f,z)]$ is the transmission matrix that contains the impulse responses between positions $u_{in}$ (at the sea surface) and positions $(x_{out}, z)$ inside the medium. Green’s functions are obtained by modeling wave propagation inside the medium with the Split Step Fourier method \cite{hardin1973application} which models wave propagation as a succession of layers of diffraction events and phase screens obtained from the velocity model:
		\begin{equation}
			\mathbf{G}_{ux}(f,z)=\prod\limits_{i=0}^{N} \left(\left(\mathbf{F}\odot\mathbf{h}\left(f,c(z_i)\right)\right)\times\mathbf{F}^\dag\right) \odot \mathbf{s}\left(f,c(z_i)\right),
		\end{equation}
		with $\odot$ the element-wise product, $z_i$ the depth of layer $i$, $c(z_i)$ the wave velocity at depth $z_i$, $\mathbf{F}$ a Fourier transform operator. $\mathbf{h}$ and $\mathbf{s}$ are the diffraction and phase screen vectors defined as:
		\begin{gather}
			\mathbf{h}\left(f, c(z)\right) = \text{exp}\left(j2\pi \delta z \sqrt{\left(f \sigma_m(z)\right)^2 - |\bm{f}_x|^2}\right),\\
			\mathbf{s}\left(f, c(z)\right) = \text{exp}\left(j2\pi \delta z f \bm{\delta \sigma}(z)\right),
		\end{gather}
		with $\bm{f}_x$, the spatial lateral frequencies written as a vector, $\sigma_m(z)$, the average slowness (inverse of velocity) at depth $z$, and $\bm{\delta \sigma}(z)  = [\delta \sigma (x,z)]$, the residual lateral slowness written as a vector, such that $\delta \sigma (x,z) = 1 / c(x,z) - \sigma_m(z)$. As a wave velocity model, we choose, for the first 5 km depth, the result of full waveform inversion \cite{marjanovic2018crustal}, and use a constant extrapolation for higher depths (Fig. S4). The obtained reflection matrices can be time-gated by summing over the frequency dimension, yielding the time-gated focused reflection matrix \cite{lambert2020reflection, touma2021distortion}:
		\begin{equation}
			\mathbf{R}_{xx}(z) = \sum\limits_f \mathbf{\widehat{R}}_{xx}(f,z).
		\end{equation}
		Such a reflection matrix contains the responses between all virtual geophones synthetized at depth $z$. Examples of such matrices are represented in Figs.~\ref{fig2}C and \ref{fig2}F. Along their diagonal lies the reflectivity image,  $I(x, z)=\text{diag}\left(\mathbf{R}_{xx}(z)\right)$, which corresponds to the cases in which the virtual source is superimposed with the virtual receiver.

		\subsection*{Matrix Imaging – Aberration Correction}
		
		The focused reflection matrices contain much more information than the reflectivity image. The width of their main diagonal is representative of the width of the focal spot (i.e. width of the virtual geophone, or local resolution). In the presence of heterogeneities, this focal spot is widened, causing a loss of resolution and distortions in the reflectivity image. We observe such a widening of the reflection matrix diagonals (Fig.~\ref{fig2}C), which indicates that wavefront distortions remain despite the use of the FWI velocity model. Such distortions can be attributed to the limited depth of the velocity model (up to 5km), its lateral resolution ($>$100m), as well as any phenomenon not encompassed by this velocity data (e.g. scattering caused by density heterogeneities). Nevertheless, the matrix imaging framework allows us to isolate the effect of the residual wave velocity anomalies through the distortion matrix concept \cite{badon2020distortion, lambert2020distortion, lambert2022ultrasound}. 
		Such wavefront distortions are estimated in the k-space similarly to \cite{touma2023imaging}.
		For each depth $z$, we project the time gated focused reflection matrix $\mathbf{R}_{xx}(z)$ to the k-space basis ($\textbf{k}$) at output using the Fourier transform operator $\mathbf{F}$:
		\begin{equation}
			\mathbf{R}_{xk}(z) = \mathbf{R}_{xx}(z) \times \mathbf{F}.
		\end{equation}
		Note that this operation can be done equivalently at input, multiplying the reflection matrix on the left.
		This dual reflection matrix contains the wave-fronts associated with virtual sources corresponding to the focusing points $(x_{in},z)$ in the k-space. Those wave-fronts correspond to the sum of a geometric parabolic component related to the position of each focusing point $(x_{in},z)$ and of a distorted component resulting from the mismatch between the propagation model and the real speed-of-sound distribution.
		
		To isolate this distorted component, we subtract the geometric component that would be ideally obtained in absence of aberrations. The result is the so-called distortion matrix that can be expressed as follows:
		\begin{equation}
			\label{eq:distortion}
			\mathbf{D}_{xk}(z) = \mathbf{R}_{xk}(z) \odot \mathbf{F}^\ast. 
		\end{equation} 
		We then exploit the local correlation of the wave distortions (known as the memory effect in wave physics) to extract an aberration transmittance. This is done by computing the cross-correlation of the wavefield distortions over patches of size 3$\times$ 1.2 km in the lateral and vertical direction respectively, with 50\% overlapping in both directions: 
		\begin{equation}
			C(k, k', p) = \sum\limits_{z}\sum\limits_x w_p(x, z) D(x, k, z) D(x, k', z)^\ast, 
		\end{equation}
		with $p$ the considered patch and $w_p(x, z)$ a windowing function restricting the domain to $p$.
		
		The obtained cross-correlation can be stacked into a correlation matrix $\mathbf{C}_{kk}(p)=[ C(k, k', p)]$.
		An iterative phase reversal algorithm \cite{bureau2023three} is then applied to this matrix to extract an estimation of the aberration phase transmittance over each patch $p$:
		\begin{equation}
			\left\lbrace \begin{array}{l}
				\bm{\phi}_0 = \mathbf{0},\\
				\bm{\phi}_{n+1} = \text{arg}\left\lbrace \mathbf{C}_{kk}(z) \times \text{exp}\left(j  \bm{\phi}_n \right) \right\rbrace.
			\end{array}\right.
		\end{equation}
		in which $\text{arg}$ is the phase operator. This process is iterated until convergence: $\phi= \lim\limits_{n \rightarrow \infty} \phi_n$. The aberrations are finally corrected by applying the phase conjugate of the aberration transmittance, $\text{exp}\left(-j  \bm{\phi} \right)$, to the distortion matrix, before projecting it back to the focused basis, yielding a corrected focused reflection matrix:
		\begin{equation}
			\mathbf{R}_{xx}(z) = \left(\mathbf{R}_{xk}(z) \odot \text{exp}\left(-j  \mathbf{\phi} \right)^T \right) \times \mathbf{F}^\dag. 
		\end{equation}
		This correction is performed two times, both in the output side and in the input side (using $\mathbf{D}_{kx}(z)$ instead of $\mathbf{D}_{xk}(z)$), to compensate for aberrations induced by wave velocity anomalies on both the reflected and incident waves. The second iteration is to ensure convergence of the entire aberration correction process. This aberration correction restores a sharp focusing (see focused reflection matrices of Fig.~\ref{fig2}F) and a corrected reflectivity image (Fig.~\ref{fig2}E) with an optimized contrast and a resolution close to the diffraction limit.

		\section*{REFERENCES AND NOTES}
		
		\bibliography{references}
		\bibliographystyle{ScienceAdvances}
		
		\newpage

		\newpage
		\section*{Supplementary Figures}
		\setcounter{figure}{0}
		\renewcommand{\thefigure}{S\arabic{figure}}
		
		\begin{figure}[ht]
			\centering
			\includegraphics[width=0.5\textwidth, trim={0mm 1mm 0mm 0mm}, clip]{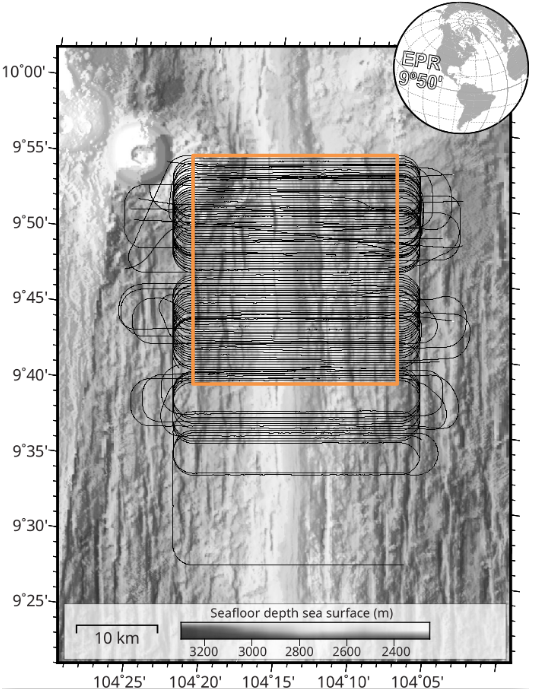}
			\caption{\label{figs1} \textbf{Survey geometry.} Bathymetry map with cross-axis ship tracks (thin black lines) displaying racetrack geometry. The orange box shows the region of the 3-D migrated seismic dataset we present here. The globe indicates the location of the East Pacific Rise (EPR) 9º50'N, marked in a black star. Modified from \cite{marjanovic2023insights}.}
		\end{figure}
		
		\begin{figure}[ht]
			\centering
			\includegraphics[width=\textwidth]{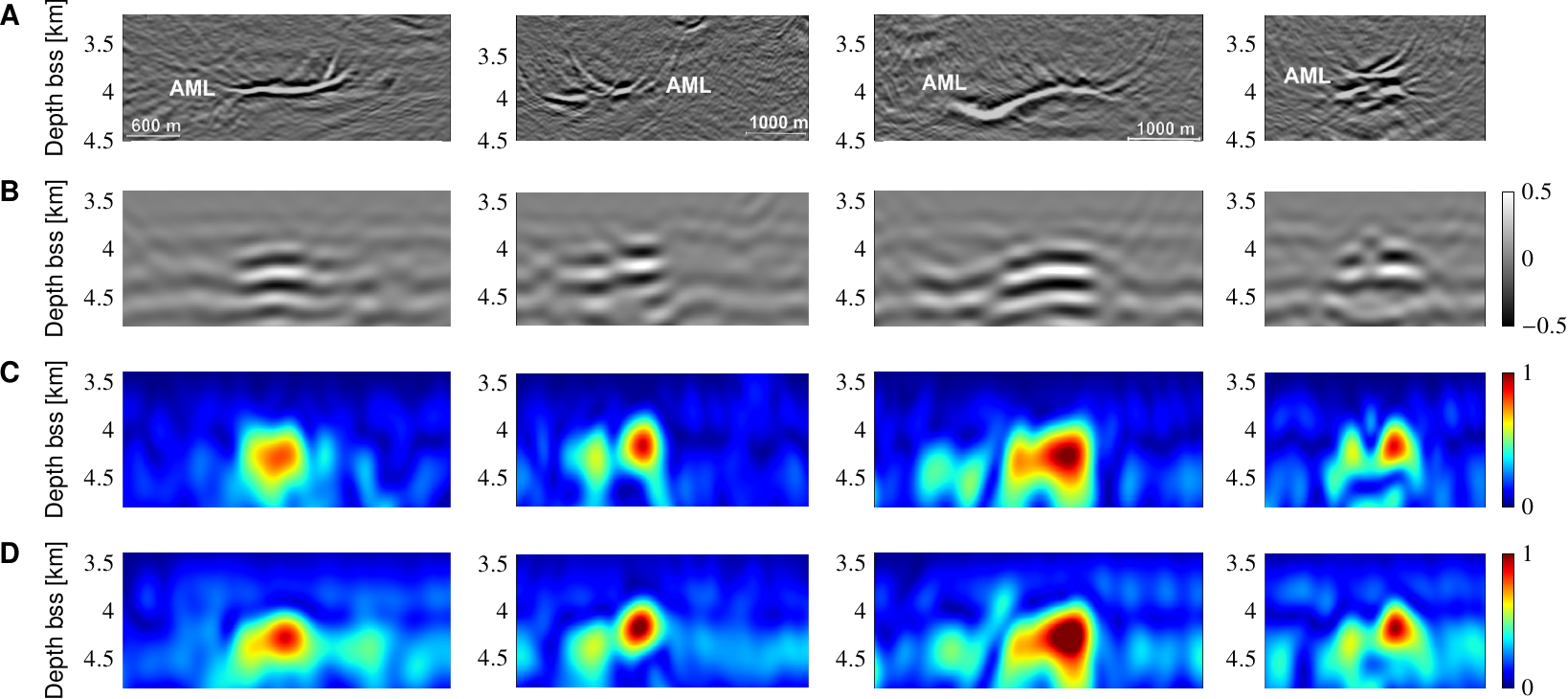}
			\caption{\label{figs2}\textbf{Resolving axial magma lenses with matrix imaging.} (\textbf{A}) Examples of high-frequency \bapt{(up to 100 Hz)} images of AML from pre-stack depth migrated volume using FWI velocity model. (\textbf{B}) Low-frequency imaging of AML with matrix imaging using FWI velocity model. (\textbf{C}) Corresponding reflectivity images obtained by taking the envelope of ({B}).  (\textbf{D}) Reflectivity images obtained after aberration correction. The corrected images (D) are quite close to the initial images (B) for the AML, which suggests that the FWI velocity model is \baptdel{correct}\bapt{sufficient} above the AML.
}
		\end{figure}
		
		\begin{figure}[ht]
			\centering
			\includegraphics[width=\textwidth]{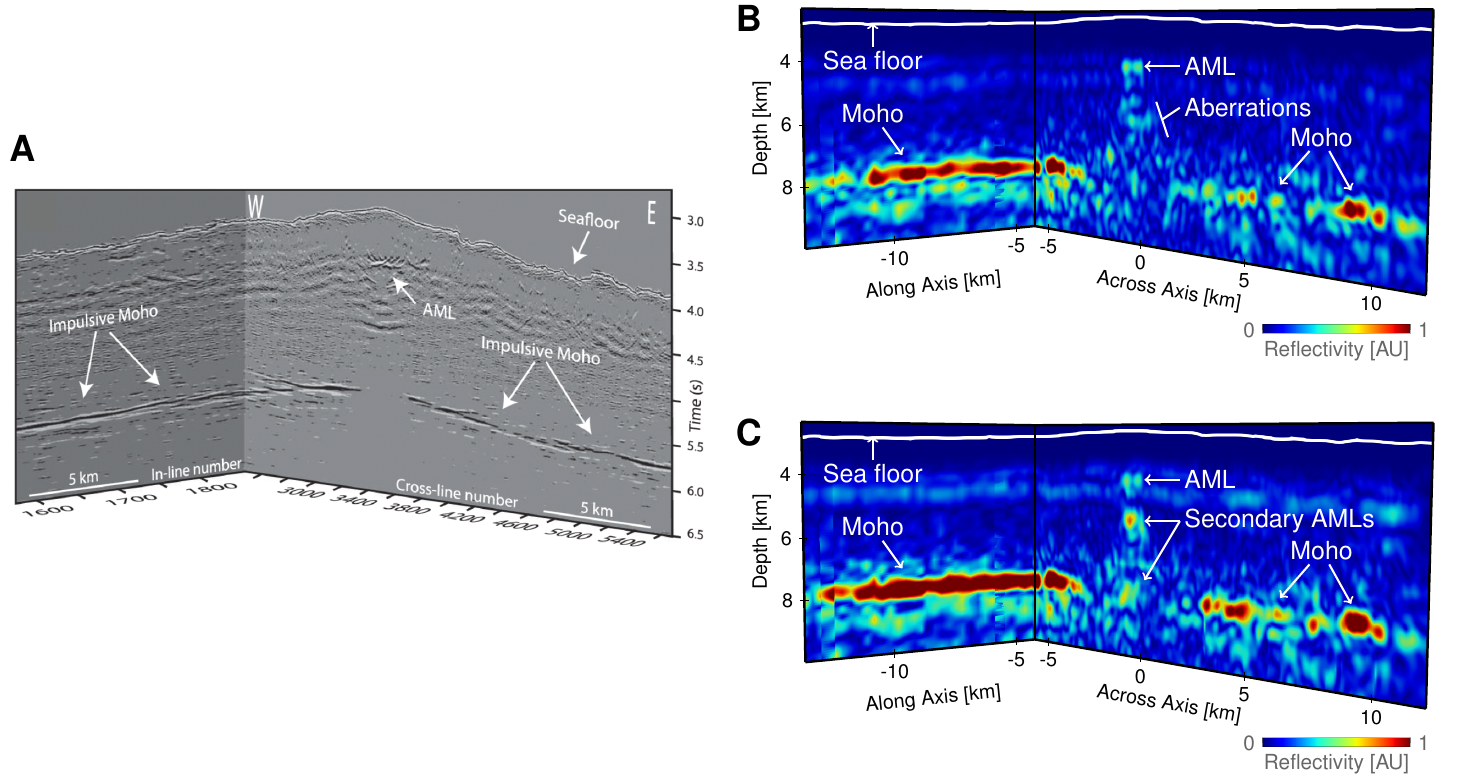}
			\caption{\label{figs3}\textbf{Imaging until the Moho with matrix imaging.} (\textbf{A}) The strong impulsive Moho reflector imaged in the 3-D post-stack Kirchhoff time migrated seismic volume, modified from \cite{aghaei2014crustal}. The E-W (inline) crosses the ridge axis at ~9º52.2’N. In addition to Moho, the presence of the reflections of the seafloor and AML beneath the ridge axis are indicated. (\textbf{B}) Moho imaged with matrix imaging, at approximately the same locations as in (A). The presence of the AML and following distorted echoes are indicated. (\textbf{C}) Same results obtained after correction, showing a stronger Moho and AML echoes as well as secondary AMLs.
}
		\end{figure}

		\begin{figure}[ht]
			\centering
			\includegraphics[width=\textwidth]{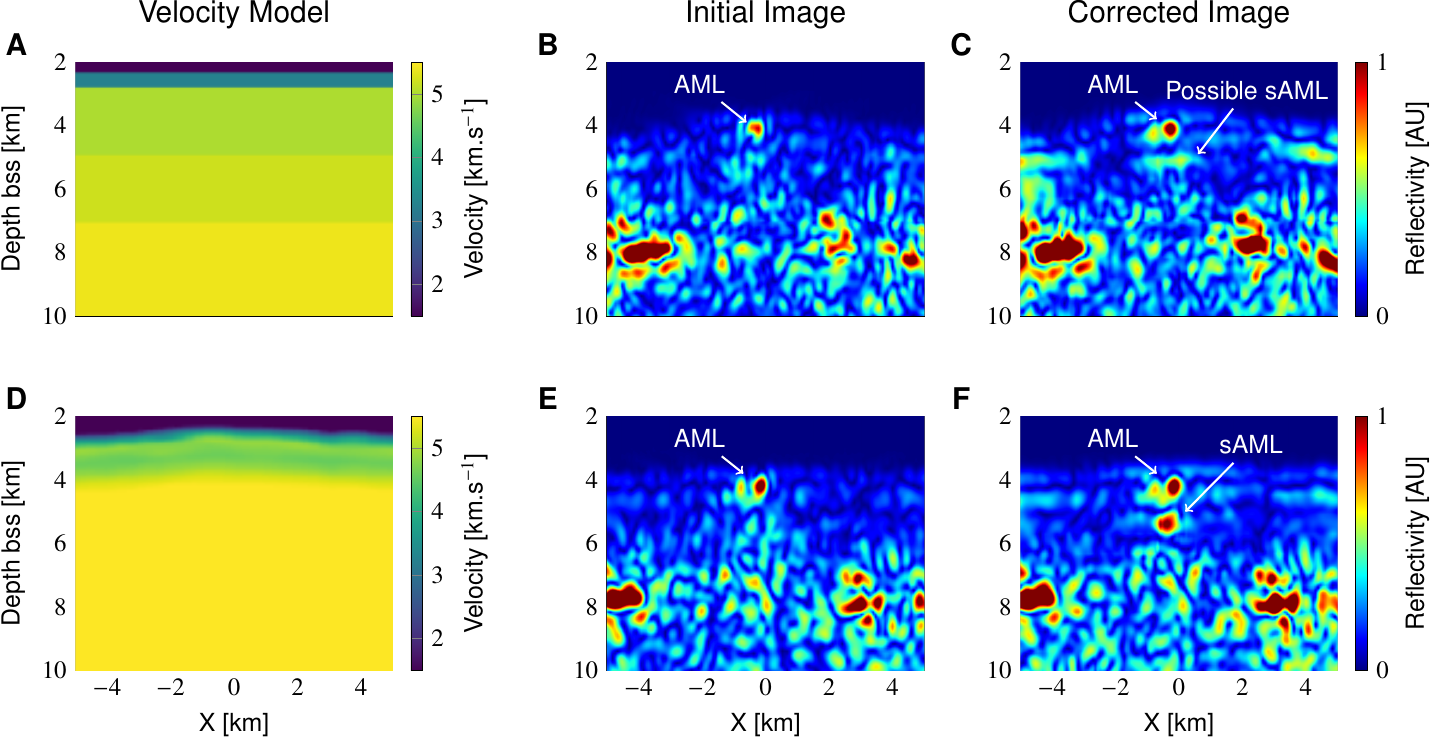}
			\caption{\label{figs4}\textbf{Combining FWI and Matrix Imaging.} (\textbf{A}) Layered velocity model. (\textbf{B}) Reflectivity image obtained with matrix imaging, using the layered velocity model in \textbf{A}. The presence of the AML is indicated. (\textbf{C}) Same results obtained after correction, showing a stronger AML echoes as well as a possible secondary AML.
			(\textbf{D}) FWI velocity model obtained in \cite{marjanovic2018crustal}, with a constant extrapolation at high depth. (\textbf{E-F}) Same as \textbf{C-D}, but using the FWI velocity model. The aberration correction performed with matrix imaging result in a clearer secondary AML, as well as a reduced level of reflectivity outside of the axis centered reservoir.
}
		\end{figure}
	\end{document}